\documentclass[aps,prx,twocolumn,groupedaddress,amsmath,amssymb]{revtex4-1}
\usepackage{graphicx,bm}

\begin{document}


\title{Solution of the multi-state voter model and application to strong neutrals in the naming game}


\author{William Pickering}
\author{Chjan Lim}
\affiliation{Department of Mathematical Sciences, Rensselaer Polytechnic Institute, 110 8th Street, Troy, New York 12180, USA}

\date{\today}
\begin{abstract}
We consider the voter model with $M$ states initially in the system. Using generating functions, we pose the spectral problem for the Markov transition matrix and solve for all eigenvalues and eigenvectors exactly. With this solution, we can find all future probability probability distributions, the expected time for the system to condense from $M$ states to $M-1$ states, the moments of consensus time, the expected local times, and the expected number of states over time. Furthermore, when the initial distribution is uniform, such as when $M=N$, we can find simplified expressions for these quantities. In particular, we show that the mean and variance of consensus time for $M=N$ is $\frac{1}{N}(N-1)^2$ and $\frac{1}{3}(\pi^2-9)(N-1)^2$ respectively.

\end{abstract}


\maketitle

\section{Introduction}

The voter model is a well studied model in social opinion dynamics \cite{liggett,liggett2,clifford,sen,castellano}.
In the binary case, each node in a network is endowed with one of two states. In a single update, a node is chosen randomly
and adopts the state of a randomly chosen neighbor. Although the model typically specifies that the states are binary,
we study the  case in which there are initially $M$ states \cite{starnini, vazquez2, volovik, castello2}. We assume $M$
can take any value from $2$ to $N$, where $N$ are the number of nodes in the network. As the system evolves, it inevitably
eliminates a state completely. That is, there will almost surely be $M-1$ distinct states in the network at some finite
future time. This process is repeated until consensus is reached and one opinion dominates the network.

The microscopic rules are unchanged by introducing more states in the multi-state extension that we consider. Other
extensions that modify the update rules by introducing more states have also been studied \cite{vazquez2, volovik,
castello2}. The two word Naming Game, for instance, is another social and linguistic model that updates with different
rules than the voter model and has noticeably distinct features \cite{baronchelli2, zhang, xie}. Multi-state extensions
allow for the consideration of neutral opinions, and the behavior of these individuals has a vital role in the resulting
features of the system. For the Naming Game, this intermediate state assumes that individuals are flippant in the sense
that they always convey an extreme opinion with equal probability. When compared with the three state voter model, the
resulting features of the model are in great contrast with those of the Naming Game \cite{zhang,lu}. The third state in
the voter model can be thought of as a strongly neutral position \cite{lim13}. In this context, neutrality itself
constitutes an opinion of its own that is independent of the other states. Moderate individuals who accept a compromise
between the two extreme beliefs while rejecting the extremes themselves would fall into this category. They would tend to
speak the moderate opinion instead of either extremes. This paper will provide much insight into the properties of social
systems with strong neutrals in contrast to the weak neutral cases that have been considered previously in the Naming
Game.

One may observe that many social discussions pose a dichotomy between opposing viewpoints. Despite the unique characteristics of each individual person, like-minded groups often form quickly and dominate the discourse. This is certainly evident when considering political parties, which show that only a few distinct opinions are expressed on a national scale. It may seem counterintuitive that these associations form so quickly when individual thought would initially provide for much disagreement. We will show that the multi-state voter model reconciles how social and political systems quickly condense into only a few dominating opinions for any value of $M$. We will show that when $M=N$, the expected time to consensus is barely larger than when $M=2$. Furthermore, it will be shown that $O(N)$ opinion states will be eliminated in $O(1)$ time. These results indicate that the opinions of individuals quickly condense together into a few dominating groups and that the system reaches consensus at a much slower rate.

We will outline a very powerful procedure that can diagonalize the transition matrix for the multi-state voter model. This procedure is a generalized application of the methods used to diagonalize the binary state voter model \cite{pickering}. The model is viewed as an urn problem in which two balls are chosen randomly and placed in the urn from which the second ball was chosen. By using generating functions, we will solve for all eigenvalues and eigenvectors of the transition matrix, which in turn allows us to easily find all future probability distributions. These techniques can be traced to the solution of the Ehrenfest urn model \cite{kac}. The multi-state voter model can be thought of as an $M$ urn system will $N$ balls distributed amongst them with the same update rules.  To solve this urn system by generating functions, we cast the spectral problem as a partial differential equation that can be solved. This the most natural approach since each independent variable in the generating function corresponds to an urn/opinion state.

The paper will be outlined as follows. In Sec. \ref{sec:model}, we clearly define the complete graph model and the notation that will be used. In Sec. \ref{sec:spectral}, we will consider the Markov transition matrix for a single update and solve for its eigenvalues and eigenvectors. We use generating functions to solve the spectral problem \cite{pickering,kac}. This solution has several applications and consequences that will be described in detail in Sec. \ref{sec:applications}. In particular, we consider the $m$ step propagator, moments of consensus time, expected local time, expected time to collapse, and exact moments of consensus time. Also in Sec. \ref{sec:applications}, we consider the case when the initial distribution is uniform, which is always the case when $M=N$. In uniform cases, the solutions we provide simplify considerably.

\section{The Multi-State Voter Model}\label{sec:model}
We assume throughout that there are $N$ nodes and that every node is connected to all other nodes. We apply the model on the complete graph primarily for analytical tractability. Each node in the network is initially endowed with one of $M$ possible opinion states denoted by $A_1,A_2,\ldots,A_M$. We define the components of $\mathbf{n}(m)$ to be total number of nodes with opinion $A_j$. The voter model prescribes the random walk for the macro-state vector $\mathbf{n}$. That is, we can write for time step $m$,

\begin{equation}
\begin{pmatrix}
n_1(m+1)\\
\vdots\\
n_M(m+1)
\end{pmatrix}
=
\begin{pmatrix}
n_1(m)\\
\vdots\\
n_M(m)
\end{pmatrix}
+
\begin{pmatrix}
\Delta n_1(m)\\
\vdots\\
\Delta n_M(m)
\end{pmatrix}.
\end{equation}
Here, $\Delta n$ contains the random nature of the walk at time step $m$. In a single update, $\Delta n_i=1$ and $\Delta n_j =-1$ for some $i$ and $j$, which implies that $\Delta\mathbf{n}=\mathbf{e}_i-\mathbf{e}_j$ for standard basis vectors $\mathbf{e}_k$. The probability that $\Delta\mathbf{n}$ takes this value is prescribed by the rules of the voter model and is given by

\begin{equation}
\Pr\{\Delta\mathbf{n}=\mathbf{e}_i-\mathbf{e}_j|\mathbf{n}(m)=\bm{\alpha}\}=\frac{\alpha_i\alpha_j}{N(N-1)}.
\end{equation}
This accounts for all lazy steps in the system as well since there is a non-zero probability that $\Delta\mathbf{n}=\mathbf{0}$.

We define the macro-state probability distribution by $a_{\bm{\alpha}}^{(m)}=Pr\{\mathbf{n}(m)=\bm{\alpha}\}$. We now define a generating function for the probability distribution of macro-states at time step $m$ as 
\begin{equation}
Q^{(m)}(\bm{x})=\sum_{|\bm{\alpha}|=N} a_{\bm{\alpha}}^{(m)}\bm{x^\alpha}.
\end{equation}
The vector power is interpreted in the sense of the multi-index notation of Laurent Schwartz\cite{john}, which shall be used extensively. This generating function allows us to very easily find a succinct expression for the Markov transition matrix for a single step of the multi voter model. The form of the generating function allows us to determine the shift and differentiation properties of $Q^{(m)}$. These properties are given by

\begin{enumerate}
\item $\frac{\alpha_i\alpha_j}{N(N-1)}a_{\bf{\alpha}}^{(m)}\longrightarrow \frac{x_ix_j}{N(N-1)}Q_{x_ix_j}^{(m)}$
\item $a_{\bm{\alpha}-\bm{e}_i+\bm{e}_j}^{(m)}\longrightarrow \frac{x_i}{x_j}Q^{(m)}$ \cite{newman,newman2,bender,pickering}.
\end{enumerate}
Using these properties, we can rewrite the spectral problem as an equivalent partial differential equation for the generating function of the macro-state probability distribution as

\begin{equation}
Q^{(m+1)}-Q^{(m)}=\sum_{i=1}^{M-1}\sum_{j=i+1}^M \frac{(x_i-x_j)^2}{N(N-1)}\frac{\partial^2Q^{(m)}}{\partial x_i\partial x_j} \label{GFpropagator}.
\end{equation}
This constitutes a transition matrix that we wish to diagonalize. Given the diagonalization of the transition matrix, we can find all future macro-state probability distributions explicitly, which yield several exact solutions. To accomplish this, we proceed to solve for all of its eigenvalues and eigenvectors.

\section{Spectral Solution}\label{sec:spectral}
We can solve the partial differential equation given in Eqn. \eqref{GFpropagator} exactly for all eigenvalues and eigenfunctions. To do this, we write the eigenvalue problem in generating function form. For eigenvalue $\lambda$ with eigenvector $\mathbf{v}$ with components $c_{\bm{\alpha}}$. Let 
\begin{equation}
G(\mathbf{x})=\sum_{|\bm{\alpha}|=N} c_{\bm{\alpha}}x^{\bm{\alpha}}
\end{equation}
be the generating function for the eigenvector $\mathbf{v}$. Furthermore, we require that each component in the vector $\bm\alpha$ is non-negative. This is because the index $\bm\alpha$ directly represents the number of individuals with each opinion type. We can rewrite the eigenvalue problem for Eqn. \eqref{GFpropagator} as 

\begin{equation}
N(N-1)(\lambda-1)G=\sum_{i=1}^{M-1}\sum_{j=i+1}^M (x_i-x_j)^2\frac{\partial^2G}{\partial x_i\partial x_j}\label{G}.
\end{equation}
We solve for both $\lambda$ and $G$ by utilizing a linear change of variables $\mathbf{x}\rightarrow\mathbf{u}$ and $G(\mathbf{x})=H(\mathbf{u})$. Since the change of variables is linear, we expect $H$ to have the same form as $G$. So, we define $b_{\bm{\alpha}}$ so that
\begin{equation}
H(\mathbf{u})=\sum_{|\bm{\alpha}|=N}b_{\bm{\alpha}}\mathbf{u}^{\bm{\alpha}}\label{Hdefn}.
\end{equation} 
The change of variables is chosen so that the resulting difference equation for $b_{\bm{\alpha}}$ is explicit. Having an explicit equation for $b_{\bm{\alpha}}$ will allow us to to find all eigenvalues and eigenvectors exactly. The change of variables that accomplishes this is given to be

\begin{align}
&u_1=x_1-x_M\\
&\;\;\;\;\vdots\\
&u_{M-1}=x_{M-1}-x_M\\
&u_M=x_M.
\end{align}

With this change of variables we can write Eqn. \eqref{G} as 

\begin{multline}
N(N-1)(\lambda-1)H=\sum_{i=1}^{M-1}\bigg[u_i^2\bigg(H_{u_iu_M}-\sum_{j=1}^{M-1}H_{u_iu_j}\bigg)\\
+\sum_{j=i+1}^{M-1}(u_i-u_j)^2H_{u_iu_j} \bigg].
\end{multline}
To simplify this equation, we use the following identity:
\begin{multline}
\sum_{i=1}^{M-1}\sum_{j=i+1}^{M-1}(u_i^2+u_j^2)H_{u_iu_j}=\\
\sum_{i=1}^{M-1}\bigg(\sum_{j=1}^{M-1}u_i^2H_{u_iu_j}-u_i^2H_{u_iu_i} \bigg)\label{H1}.
\end{multline}
The proof of this identity is given in Appendix \ref{app:identity}. Applying the identity and canceling like terms reduces Eqn. \eqref{H1} to
\begin{multline}
N(N-1)(\lambda-1)H=\sum_{i=1}^{M-1}\bigg[-u_i^2H_{u_iu_i}\\
-\sum_{j=i+1}^{M-1}2u_iu_jH_{u_iu_j}+u_i^2H_{u_iu_M} \bigg].
\end{multline}
We now rewrite this as a difference equation for the coefficients of $H$. By Eqn. \eqref{Hdefn}, we obtain

\begin{multline}
N(N-1)(\lambda-1)b_{\bm{\alpha}}=\sum_{i=1}^{M-1}\bigg[-\alpha_i(\alpha_i-1)b_{\bm{\alpha}}\\
-\sum_{j=i+1}^{M-1}2\alpha_i\alpha_jb_{\bm{\alpha}}+(\alpha_i-1)(\alpha_M+1)b_{\bm{\alpha-e_i+e_M}}\bigg]\label{H2}.
\end{multline}
Let $w(\bm\alpha)=\sum_{i=1}^{M-1}\alpha_i$. We use this to reduce Eqn. \eqref{H2} to an explicit form given by
\begin{equation}
b_{\bm\alpha}=\frac{(\alpha_M+1)\sum_{i=1}^{M-1}(\alpha_i-1)b_{\bm{\alpha-e_i+e_M}}}{N(N-1)(\lambda-1)+w(\bm\alpha)[w(\bm\alpha)-1]}\label{explicit}.
\end{equation}
Recall that each $\alpha_i\geq0$.

Observe that if Eqn. \eqref{explicit} is non-singular for every $\bm\alpha$, then every $b_{\bm\alpha}=0$. Since this corresponds to the trivial solution to the eigenvalue problem, we discard these solutions. Requiring a singularity in Eqn. \eqref{explicit} implies that the eigenvalues for the transition matrix of the $M$ state voter model are

\begin{equation}
\lambda_{w(\bm\alpha)}=1-\frac{w(\bm\alpha)[w(\bm\alpha)-1]}{N(N-1)}\label{eigenvalues}.
\end{equation}
Since many values of $\bm\alpha$ will yield the same value for $w(\bm\alpha)$, it is clear that there will be many repeated eigenvalues. Given that $w(\bm\alpha)$ ranges from $0$ to $N$, the set of eigenvalues is the same for all values of $M$. However, the multiplicities of each eigenvalue will vary with $M$.

The components of the eigenvectors can be found by transforming back from $H(\bm u)$ to $G(\bm x)$. This will yield a relationship between $c_{\bm\alpha}$ and $b_{\alpha}$. Using generating function techniques, this relationship is given by

\begin{equation}
c_{\bm\alpha}=\sum_{|\bm{\beta}|=N}b_{\bm\beta}(-1)^{\alpha_M-\beta_M}\prod_{i=1}^{M-1} {\beta_i\choose \alpha_i} .\label{eigenvector_components}
\end{equation}
Here, $\bm\beta$ is a multi-index that has $M$ non-negative components, similar to $\bm\alpha$. The mathematical derivation of Eqn. \eqref{eigenvector_components} is given in Appendix \ref{app:eigenvector}.

\section{Applications}\label{sec:applications}
With the solution to the spectral problem available, we can exactly calculate several quantities and estimate others. Below we define and calculate the expected collapse times, the moments of consensus time, the expected local times, and the expected number of states over time. We also consider the special case when $M=N$, which is when all nodes begin with a distinct opinion. The connection between the spectral problem and these exact solutions is as follows. With all eigenvalues and eigenvectors, we can explicitly diagonalize the Markov transition matrix for the macrostates of the system. The probability of achieving each macrostate governs each of the following quantities, so having an exact $m$-step propagator allows us to exactly calculate these solutions.

\subsection{Moments of Collapse Times}\label{collapse_times}
The collapse times, $\tau_k$, are the amount of scaled time, $m/N$, until a state is eliminated from the system. That is, if there are $k$ states in the system, then the collapse time is the time until only $k-1$ states are present. Once eliminated, a state can never be reintroduced in the system. This process is repeated until each individual adopts a single consensus state.

Here, we estimate the collapse times for each $k$. We do this by estimating the probability that $k$ have survived by at time $m$. We call this the survival probability and denote it by $S_k(m)$. Given that the system is martingale and that the solution of the spectral problem is known, the survival probability for $M$ states can be bounded as follows: 

\begin{equation}
S_k(m)=O(\lambda_k^m).\label{survival_estimate}
\end{equation}
One can verify that the dominant eigenvalues are $\lambda_k$ by considering the eigenvector $c_{\bm\alpha}=1$ and assuming that the index $\bm\alpha$ corresponds to macrostates where $k$ states have survived. This eigenvector provides a uniform upper bound for the survival probability the system.

To find the moments of the time to collapse, we use this to estimate the probability that the system collapses at time $m$. This is equal to the difference between the survival probabilities from time step $m-1$ to $m$. That is, by Eqn. \eqref{survival_estimate}, the probability of collapse at time $m$ is $O[\lambda_k^{m-1}(1-\lambda_k)]$. Therefore, we can write the $pth$ moments of collapse time as 

\begin{align}
E[\tau_k^p]&=(1-\lambda_k)\sum_{m=1}^\infty O\left(\lambda_k^m \left(\frac{m}{N}\right)^p\right)\\
&=O\left\{ p!\left[\frac{N-1}{k(k-1)}\right]^p\right\}.\label{collapse}
\end{align}
The estimate varies for various initial conditions, but is asymptotically correct for all $N$, $k$, and $p$. We will make use of this result when determining the asymptotic behavior of the moments of consensus time and the expected number of states over time.

\subsection{Moments of Consensus Time}\label{moments_of_consensus_time}
The consensus time, $\tau$, is the amount of scaled time until every individual adopts a single opinion. All consensus states are absorbing, so once this state has been achieved, all dynamics in the system halt. We can use the solution to the spectral problem to find all moments of the consensus time. Furthermore, we will use estimates to find the asymptotic behavior for large $N$.

To find the consensus time, we define $l^{(m)}$ to be the probability that the system reaches consensus at time $m$. This is equal to the probability that the system has only one individual that has a different state than all of the others and then adopts the majority opinion. Therefore, we write

\begin{equation}
l^{(m)}=\frac{1}{N}\sum_{i=1}^M\mathop{\sum_{j=1}^M}_{j\not= i} a_{\bm{e_i}+(N-1)\bm{e_j}}^{(m)}.
\end{equation}
Given the solution of the spectral problem, we can represent the macrostate probability as

\begin{equation}
a_{\bm{\alpha}}^{(m)}=\sum_{\bm{\beta}} d_{\bm{\beta}}\lambda_{\bm\beta}^m[\mathbf{v}_{\bm{\beta}}]_{\bm\alpha},\label{diagonalization}
\end{equation}
where $d_{\bm\alpha}$ is the initial distribution expressed in the eigenbasis. Let 
\begin{equation}
s_{\bm\beta}=\sum_{i=1}^M\mathop{\sum_{j=1}^M}_{j\not= i}d_{\bm{\beta}}[\mathbf{v}_{\bm{\beta}}]_{\bm{e_i}+(N-1)\bm{e_j}}.
\end{equation}
With this, the moments of the consensus time are given by

\begin{align}
E[\tau^p]&=\sum_{m=1}^\infty l^{(m)}m^p\\
&=\frac{1}{N}\sum_{m=1}^\infty \sum_{i=1}^M\mathop{\sum_{j=1}^M}_{j\not= i} \sum_{\bm{\beta}} d_{\bm{\beta}}\lambda_{\bm\beta}^m[\mathbf{v}_{\bm{\beta}}]_{\bm{e_i}+(N-1)\bm{e_j}} m^p\\
&\sim \frac{1}{N} \sum_{\bm{\beta}}\frac{p!s_{\bm\beta}}{(1-\lambda_{\bm\beta})^{p+1}}.\label{consensus_moments}
\end{align}

This is an exact solution for the moments of consensus time. Note that the quantity $s_{\bm\beta}$ depends on the initial distribution through $d_{\bm\alpha}$. The eigenvectors, $\bm v_{\bm\beta}$ can be determined component-wise by Eqn. \eqref{eigenvector_components}.

The formula given in Eqn. \eqref{consensus_moments} is exact for all $N$, $p$, and initial conditions. We now extract asymptotic information about the moments of consensus time. In particular, observe that the consensus time is the sum of all collapse times. Therefore, 

\begin{align}
\tau^p&=\left(\sum_{k=2}^M \tau_k\right)^p\\
&=\sum_{|\bm\gamma|=p} {p \choose \bm\gamma}\prod_{k=2}^M \tau_k^{\gamma_k}.
\end{align} 
Here, $\bm\gamma$ is a vector with components $\gamma_2\ldots\gamma_M$. The multi-index notation is used to denote the multinomial coefficients as well. Also, the collapse times are independent random variables. So, when taking the expected value of $\tau^p$, we obtain

\begin{equation}
E[\tau^p]=\sum_{|\bm\gamma|=p} {p \choose \bm\gamma}\prod_{k=2}^M E[\tau_k^{\gamma_k}].\label{consensus_estimate}
\end{equation}
Using the estimate in Eqn. \eqref{collapse}, this becomes
\begin{align}
E[\tau^p]&=\sum_{|\bm\gamma|=p} {p \choose \bm\gamma}\prod_{k=2}^M O\left\{\frac{\gamma_k!}{[N(1-\lambda_k)]^{\gamma_k}} \right\}\\
&=p!N^{-p}\sum_{|\bm\gamma|=p}\prod_{k=2}^M O\left[\frac{1}{(1-\lambda_k)^{\gamma_k}}\right]\\
&=O\left(p!N^p\sum_{|\bm\gamma|=p}\prod_{k=2}^M \left[\frac{1}{k(k-1)}\right]^{\gamma_k}\right)\\
&=O\left(p!N^p2^{-p}\sum_{|\bm\gamma|=p}\prod_{k=3}^M \left[\frac{2}{k(k-1)}\right]^{\gamma_k}\right).
\end{align}

We take the big-$O$ outside of the product because as the system evolves, the macro-state probability distribution tends to the uniform distribution, which corresponds to the dominant eigenvalue in the system. This means that the estimate given in Eqn. \eqref{collapse} without the big-$O$ is the exact solution. Furthermore, the moments are bounded by the dynamics when $M=N$, which examined in Sec. \ref{M_equals_N}. The initial condition also may provide further dependence on $M$, however this dependence is bounded, which does not affect the validity of the result.

Let
\begin{equation}
\eta(M,p)=\sum_{|\bm\gamma|=p}\prod_{k=3}^M \left[\frac{2}{k(k-1)}\right]^{\gamma_k}\label{eta_fn}
\end{equation}
with
\begin{equation}
\bm{\gamma}=(\gamma_2,\gamma_3,\ldots,\gamma_M).
\end{equation}
We can therefore write

\begin{equation}
E[\tau^p]=O[p!N^p2^{-p}\eta(M,p)]\label{general_consensus_estimate}
\end{equation}

We will now provide some of the fundamental properties of $\eta(M,p)$. Intuitively, the meaning of $\eta(M,p)$ is the correction made to the estimate by changing $M$. First, we show that $\eta(M,p)$ is bounded above:
\begin{align}
\eta(M,p)&\leq \sum_{0\leq\bm{\gamma}\leq p}\prod_{k=3}^{M}\left[\frac{2}{k(k-1)}\right]^{\gamma_k}\\
&=\prod_{k=3}^M \frac{1-\left[\frac{2}{k(k-1)}\right]^{p+1}}{1-\left[\frac{2}{k(k-1)}\right]}\\
&\leq \prod_{k=3}^M \frac{k(k-1)}{k(k-1)-2}\\
&= 3\left(\frac{M-1}{M+1}\right)\label{consensus_bound}
\end{align}
This result shows that the moments of consensus time can be estimated uniformly in $M$ by $O(p!N^p2^{-p})$. The dependence on $M$ affects an $O(1)$ factor of the moments of consensus time that one may not wish to casually ignore. For instance, when $M=2$, we have $\eta(2,p)=1$ whereas $\eta(M,p)\leq 3$ for large $M$. This suggests that the uniform estimate can be up to three times as high as the exact solution as $M$ changes. Furthermore, for fixed $M$ and as $p\rightarrow\infty$, the estimate given by Eq. \eqref{consensus_bound} is exact. Therefore, we have that 
\begin{equation}
\eta(M,\infty)=3\left(\frac{M-1}{M+1}\right).
\end{equation}

\begin{figure}[h!]
\includegraphics[scale=0.5]{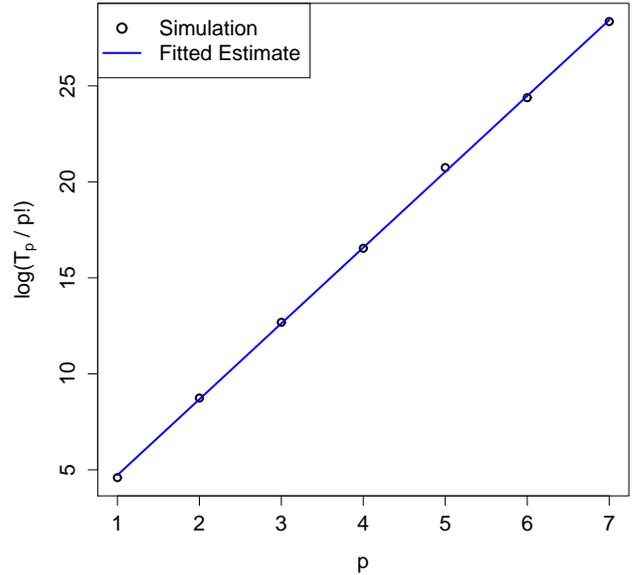}
\caption{Simulation of the voter model for $N=100$ and $M=50$ is plotted with $p$. For each $p$, the simulation is averaged over $1,000$ runs. Since $\eta$ is bounded, Eqn. \eqref{general_consensus_estimate} predicts a linear relationship with $p$ with slope $\log(100)-\log(2)\approx3.912$. The best fit line for the data is given, which has slope 3.951.}
\end{figure}

\begin{figure}[h!]
\includegraphics[scale=0.5]{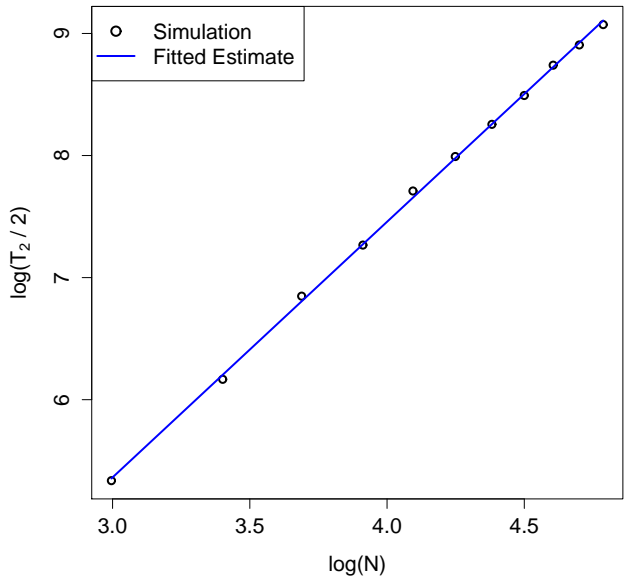}
\caption{Simulation of the voter model is plotted with $\log N$. Data is averaged over $2,000$ runs with $p=2$, $M=20$. Eqn. \eqref{general_consensus_estimate} predicts a linear relationship between the second moment and $\log N$ with a slope of $2$. The best fit line for the data is given, which has a slope of $2.094$.}
\end{figure}

For the first and second moments, evaluations of $\eta(M,p)$ are also given:
\begin{align}
\eta(M,1)&=2\left(1-\frac{1}{M}\right)\label{eta_p1}\\
\eta(M,2)&=\frac{2}{3}(\pi^2-9)+2\left(1-\frac{1}{M}\right)^2-\frac{2}{3M^3}+O\left(M^{-4}\right).\label{eta_p2}
\end{align}
Eqn. \eqref{eta_p1} shows that the expected consensus time is always $O(N)$ regardless of the number of initial states $M$. We use these particular cases will be used explicitly in Sec. \ref{M_equals_N} when studying the exact solutions of the moments of the consensus.

For small $M$, $\eta(M,p)$ can be easily calculated exactly. For $M=2,3,4$, we have

\begin{align}
\eta(2,p)&=1\\
\eta(3,p)&=\frac{3}{2}-\frac{3^{-p}}{2}\\
\eta(4,p)&=\frac{9}{5}-3^{-p}+\frac{6^{-p}}{5}.
\end{align}
Note that as $p\rightarrow\infty$, these solutions for $\eta(M,p)$ exactly match the upper bound given in Eqn. \eqref{consensus_bound}. The dependence on $p$ for each $\eta(M,p)$ always takes the form of an exponential attraction to the upper bound in Eqn. \eqref{consensus_bound}.

\subsection{Expected Local Times}
The local time is defined as the amount of scaled time, $m/N$, spent at each macrostate $\bm n$ prior to consensus. If $M_{\bm \alpha}(m)$ is the number of times state $\bm n=\bm\alpha$ has been visited by time $m$, then one can construct a random walk model for each $M_{\bm\alpha}$. That is, we write
\begin{equation}
M_{\bm\alpha}(m+1)=M_{\bm\alpha}(m)+\Delta M_{\bm\alpha}(m). \label{local_RW}
\end{equation}
The expected local time, therefore, is $E[M_{\bm\alpha}(\infty)]$ with this notation. Taking the expected value of Eqn. \eqref{local_RW} and summing from $m=0$ to $m=\infty$, we get
\begin{equation}
E[M_{\bm\alpha}(\infty)]=E[M_{\bm\alpha}(0)]+\sum_{m=0}^{\infty} E[\Delta M_{\alpha}(m)].
\end{equation}
Now, $M_{\bm\alpha}(0)=1$ if $\bm n(0)=\bm\alpha$ and equals $0$ otherwise. Therefore, $E[M_{\bm\alpha}(0)]=a_{\bm\alpha}^{(0)}$, which is given by the initial condition. Similarly, $\Delta M_{\alpha}(m)=1$ if $\bm n (m+1)=\bm\alpha$ and equals $0$ otherwise. The probability that $\Delta M_{\alpha}(m)=1$ is $a_{\alpha}^{(m+1)}$. So, the local time for state $\bm n=\bm\alpha$ is
\begin{align}
E[M_{\bm\alpha}(\infty)]&=a_{\bm\alpha}^{(0)}+\sum_{m=0}^{\infty} a_{\bm\alpha}^{(m+1)}\\
&=\sum_{m=0}^{\infty} a_{\bm\alpha}^{(m)}
\end{align}
We use the diagonalization given in Eqn. \eqref{diagonalization} to compute this. We ignore the terms that have eigenvalue $1$ however because these correspond to consensus states. We only consider non-absorbing states when considering local time. Let $E[\bm M]$ take components $E[M_{\bm\alpha}(\infty)]$. Therefore, the local time reduces to

\begin{equation}
E[\bm M]=\mathop{\sum_{\bm\beta}}_{\lambda_{\bm\beta}\not=1}\frac{d_{\bm\beta}\bm v_{\bm\beta}}{1-\lambda_{\bm\beta}}.
\end{equation}
The components of $\bm M$ that correspond to consensus states are meaningless, as it is understood that when the system enters a consensus state, the dynamics halt entirely. The other components are exactly equal to the expected local time for their respective macrostates.

\begin{figure}[h!]
\includegraphics[scale=0.5]{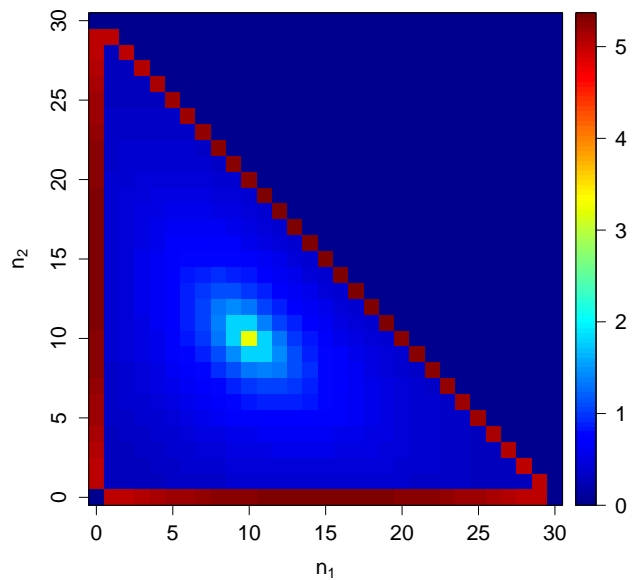}
\caption{Example of expected local times for $N=30$, and $M=3$. The initial condition is $n_1(0)=n_2(0)=10$. Most of the time is spent on the boundary where one of the states had been eliminated. Each macrostate on the boundary has nearly equal local time.}
\end{figure}

\subsection{Expected States over Time}\label{states_over_time}
Given the solution to the spectral problem and the expected collapse times, finding the expected number of states over time, $s(t)$, is straightforward. To do this, we sum the collapse times, $\tau_k$, from $k=s+1$ to $k=M$. Using Eqn. \eqref{collapse}, we show that the time for $s$ states to exist in the system is
\begin{align}
t&=\sum_{k=s+1}^{M} O\left(\frac{N}{k(k-1)}\right)\\
&=O\left[N\left(\frac{1}{s}-\frac{1}{M}\right)\right]
\end{align}  
Here, $t$ is interpreted as the scaled time $m/N$. Solving for $S$ shows that the expected number of states as a function of time is
\begin{equation}
s(t)=\left(\frac{1}{M}+\frac{ct}{N}\right)^{-1}
\end{equation}
for a constant rate $c$. This result is in agreement with the literature regarding the multi-state voter model \cite{starnini}.

\subsection{Ordering Dynamics for Uniform Distributions and $M=N$}\label{M_equals_N}
While the above solutions hold for all $M$, $N$, $p$, and initial condition $a_{\bm\alpha}^{(0)}$, the ordering dynamics of the model reduce significantly in the special case where the initial condition is uniformly distributed. This is because the uniform distribution is also an eigenvector for all $M$. The eigenvalue that corresponds to this eigenvector is $\lambda_k$ when there are $k$ distinct opinions in the system. Therefore, the diagonalization reduces considerably, which allows us to find simplified expressions for the above quantities.

A special case of a uniformly distributed initial condition is when $M=N$. This is when each individual adopts a unique, personal opinion state prior to global discussion. In this case, there is only one possible initial condition, and therefore uniformly distributed. Also, notice that in the next iteration of the model, one state will have been eliminated with probability 1. In this time step, two individuals will have the same state while the others possess distinct states. The probability distribution at this time step is uniform (constant). That is, each state is equally likely to have the two individuals than any other state during the first time step.

We can compute $\tau_k$ exactly for each $k$ for the uniform case. Because the probability distribution of the macrostates is an eigenvector, the estimates we calculated above are exact. In particular, the survival probability is given to be $S_k(m)=\lambda_k^m$. Thus, making this substitution into the derivation given in Sec. \ref{collapse_times}, we find that the expected time to collapse from $k$ states to $k-1$ states is given to be exactly
\begin{equation}
E[\tau_k]=\frac{N-1}{k(k-1)}\label{collapse_time_moments}
\end{equation}
We now use this to find the exact number of states over time. Recall from Sec. \ref{states_over_time} that the time to achieve $s$ states is the sum of collapse times from $k=s+1$ to $k=M$. Therefore, we obtain

\begin{align}
t&=\sum_{k=s+1}^M \frac{N-1}{k(k-1)}\\
&=(N-1)\left(\frac{1}{s}-\frac{1}{M}\right)
\end{align}
Therefore, the expected number of states is given to be
\begin{equation}
s(t)=\left(\frac{t}{N-1}+\frac{1}{M}\right)^{-1}
\end{equation}
When $M=O(N)$, this shows that $O(N)$ states will be eliminated in $O(1)$ time. For example, take $t=1$ and $M=N$ and observe that $s(1)\sim N/2$. This shows that the system retains only half of its initial number of states at $t=1$, which corresponds to a sweep of nodes in the network. For any $t=O(1)$, we find that only a fraction of the initial number of states remain, so $O(N)$ states were eliminated in this time. This shows that these systems quickly converge to $O(1)$ states relative to the consensus time.

Taking $s=1$, the resulting value of $t$ is the expected time to reach consensus. Doing so shows that the expected time to consensus is

\begin{equation}
E[\tau]=\frac{(N-1)(M-1)}{M}.\label{exact_consensus_time}
\end{equation}
For $M=N$, this result also shows that the expected number of interactions between individuals until consensus is reached is exactly $(N-1)^2$ and that the consensus time as close to $N$. When $M=2$, the expected consensus time is at most $N\log 2$ \cite{sood}, which is not much less than the $M=N$ case.

We also expand the methods in Sec. \ref{moments_of_consensus_time} to find all moments of consensus time. By combining the observation in Eqn. \eqref{consensus_estimate} with Eqn. \eqref{collapse_time_moments}, we can find that
\begin{equation}
E[\tau^p]=p!(N-1)^p2^{-p}\eta(M,p).\label{exact_consensus_moments}
\end{equation}
By using Eqn. \eqref{eta_p1}, note that the expected time to consensus given in Eqn. \eqref{exact_consensus_time} agrees with this result.

\begin{figure}[h!]
\includegraphics[scale=0.5]{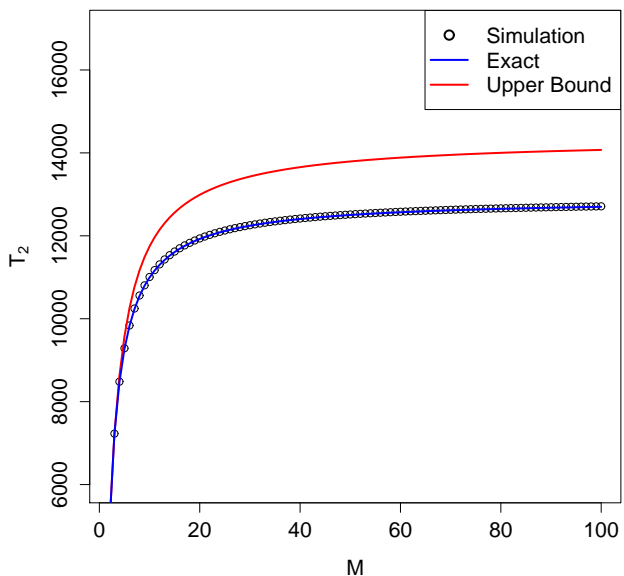}
\caption{Simulation data for the second moment of consensus time, $T_2$, with $N=100$ over $30,000$ runs is plotted as black circles. The exact solution given in Eqn. \eqref{exact_second_moment} is given as the blue curve. In addition, the upper bound found by applying Eqn. \eqref{consensus_bound}  to Eqn. \eqref{exact_consensus_moments} is plotted in red. The upper bound overestimates the data by a factor of $1.107007$ at most.}\label{M_consensus_fig}
\end{figure}

To find the second moment of consensus time, we apply Eqn. \eqref{eta_p2} to Eqn. \eqref{exact_consensus_moments} to show that

\begin{multline}
E[\tau^2]=(N-1)^2\Bigg[\frac{1}{3}(\pi^2-9)+\bigg(1-\frac{1}{M}\bigg)^2\\
-\frac{1}{3M^3}+O\bigg(\frac{1}{M^4}\bigg)\Bigg].\label{exact_second_moment}
\end{multline}
Fig. \ref{M_consensus_fig} features this result. We can also use this to find the variance of the consensus time. We combine Eqn. \eqref{exact_second_moment} with the the $p=1$ case to find that
\begin{align}
Var(\tau)=(N-1)^2\Bigg[\frac{1}{3}(\pi^2-9)-\frac{1}{3M^3}+O\bigg(\frac{1}{M^4}\bigg)\Bigg].
\end{align}
This shows that the variance of the of the consensus time for uniform distributions does not change much with $M$. Furthermore, taking $M=N$, the first term in the expansion makes for a good estimate, with higher order terms being $O(N^{-1})$. So, for $M=N$, we have $Var(\tau)\sim \frac{1}{3}(\pi^2-9)(N-1)^2$.

\section{Conclusions}\label{sec:conclusions}
The discussions above have shown many solutions to the voter model. Some of these solutions, such as the expected time to consensus and the expected states over time \cite{starnini}, confirm existing results about the ordering dynamics found by other techniques. The other solutions require the above spectral analysis to address fully. In particular, we found all moments of collapse time, all expected local times, and provided closed form expressions for all moments of the consensus time for uniform initial distributions. We then used this to find the mean and variance of the consensus time and showed that the variance converges cubicly in $M$.

The results and methods found here show that there is potential to solve other general opinion problems. Similar multi-state models that may be solvable by these techniques include the multi-allelic Moran model and the diploid Moran model of Genetic Drift \cite{ewens,moran,blythe}. The $K$-word Naming Game is one such model that has many more than $2$ states. The Naming Game with $K$-words has $2^K-1$ words, which makes numerical solution of the ODE dynamics difficult \cite{waagen}. The methods and results given here are purely analytical and therefore can circumvent those numerical challenges. Compared to the $M=3$ case of the voter model, the Naming Game has a much smaller consensus time of $O(\log N)$ \cite{baronchelli2,castello} compared to the $O(N)$ time of the 3-voter model. The variance of the consensus time for the voter model is about $6.944\times 10^{4}$ for $N=500$, which is lower than for the Naming Game, whose variance is around $1.1\times 10^7$ \cite{zhang_thesis} with committed fraction of $20\%$. Even though the Naming Game appears to give a higher variance, the result also shows that it is linear with $N$, whereas for the 3-voter, we showed that it is quadratic. So, for $N$ large enough, the variance of consensus time for the 3-voter model will eventually become much larger. 

\begin{acknowledgments}
This work was supported in part by the Army Research
Office Grant No. W911NF-09-1-0254 and W911NF-12-
1-467 0546. The views and conclusions contained in
this document are those of the authors and should not
be interpreted as representing the official policies, either
expressed or implied, of the Army Research Office or the
U.S. Government.
\end{acknowledgments}
\appendix
\section{Proof of Identity}\label{app:identity}
Here we prove the identity given in Eqn. \eqref{H1} that was utilized to solve for all eigenvalues and eigenvectors of the multi-state voter model. To begin, consider
\begin{equation}
\sum_{i=1}^{M-1}\sum_{j=1}^{M-1} u_i^2H_{u_iu_j}=\sum_{i=1}^{M-1}\left(\sum_{j=1}^iu_i^2H_{u_iu_j}+\sum_{j=i+1}^{M-1}u_i^2H_{u_iu_j}\right).
\end{equation}
For the first double sum on the right hand side, we interchange the sums. Because the sums are dependent, we obtain
\begin{equation}
\sum_{i=1}^{M-1}\sum_{j=1}^{M-1} u_i^2H_{u_iu_j}=\sum_{j=1}^{M-1}\sum_{i=j}^{M-1}u_i^2H_{u_iu_j}+\sum_{i=1}^{M-1}\sum_{j=i+1}^{M-1}u_i^2H_{u_iu_j}.
\end{equation} 
We relabel $i\leftrightarrow j$ in the first sum on the right hand side and separate the $i=j$ term to obtain.
\begin{multline}
\sum_{i=1}^{M-1}\sum_{j=1}^{M-1} u_i^2H_{u_iu_j}=\sum_{i=1}^{M-1}\left(u_i^2H_{u_iu_i}+\sum_{j=i+1}^{M-1}u_j^2H_{u_iu_j}\right)\\
+\sum_{i=1}^{M-1}\sum_{j=i+1}^{M-1}u_i^2H_{u_iu_j}.
\end{multline} 
Rearranging terms in this equation shows that
\begin{multline}
\sum_{j=i+1}^{M-1}(u_i^2+u_j^2)H_{u_iu_j}=\sum_{i=1}^{M-1}\left(\sum_{j=1}^{M-1} u_i^2H_{u_iu_j}-u_i^2H_{u_iu_i}\right).
\end{multline}
This concludes the proof of the identity of Eqn. \eqref{H1}.

\section{Calculation of Eigenvector Components}\label{app:eigenvector}
In this section, we utilize generating function techniques to relate $b_{\bm\alpha}$ to $c_{\bm\alpha}$. One strategy is to substitute $\bm u\rightarrow \bm x$ into the definition of $H(\bm u)$ and combine all terms together. By definition, this must equal $G(\bm x)$ and so the resulting coefficients must be $c_{\bm\alpha}$. For large $M$, this becomes cumbersome, so we propose a more general means of finding the relationship for any $M$ using differentiation properties of the generating functions. In particular, for multi-index derivative operator $D^{\bm\alpha}$ where $|\bm\alpha|=N$, we have
\begin{align}
D^{\bm\alpha}G(x)=\sum_{|\bm\beta|=N}c_{\bm\beta}(\bm\beta)_{\bm\alpha}\bm x^{\bm\beta-\bm\alpha}.
\end{align}
Here, $(\bm\beta)_{\bm\alpha}$ is the multi-index Pochhammer symbol, which is defined by $(\beta_1)_{\alpha_1}\ldots(\beta_M)_{\alpha_M}$. Because we defined $|\bm\alpha|=N$, the only term in the sum that is non-zero is when $\bm\beta=\bm\alpha$. Therefore, $D^{\bm\alpha}G=\bm\alpha!c_{\bm\alpha}$, where $\bm\alpha!=\alpha_1!\ldots\alpha_M!$. With this observation, we use the definition of $H$ to obtain

\begin{align}
G(\bm x)&=H(\bm u(\bm x))\\
&=\sum_{|\bm\beta|=N} b_{\bm\beta}\left[\prod_{i=1}^{M-1} (x_i-x_M)^{\beta_i}\right]x_M^{\beta_M}\\
&=\sum_{|\bm\beta|=N} b_{\bm\beta}\left[\prod_{i=1}^{M-1} \sum_{\gamma_i=0}^{\beta_i}{\beta_i\choose \gamma_i}(-1)^{\beta_i-\gamma_i}x_i^{\gamma_i}x_M^{\beta_i-\gamma_i}\right]x_M^{\beta_M}.
\end{align}
Simplifying the expression on the right side gives
\begin{multline}
G(\bm x)=\sum_{|\bm\beta|=N}b_{\bm\beta}\sum_{0\leq\bm\gamma\leq\bm\beta}\left[\prod_{i=1}^{M-1}{\beta_i\choose \gamma_i}x_i^{\gamma_i}\right]\times\\
(-1)^{N-\beta_M-|\bm\gamma|}x_M^{N-|\bm\gamma|}.
\end{multline}
Now we take $D^{\bm\alpha}$ of this equation for $|\bm\alpha|=N$. We found that on the left side, we get $\bm\alpha! c_{\bm\alpha}$. Therefore, we obtain
\begin{align}
\bm\alpha!c_{\bm\alpha}=\sum_{|\bm\beta|=N}b_{\bm\beta}\left[\prod_{i=1}^{M-1}{\beta_i\choose \gamma_i}\alpha_i!\right](-1)^{\alpha_M-\beta_M}\alpha_M!.
\end{align}
Therefore, we find that
\begin{equation}
c_{\bm\alpha}=\sum_{|\bm\beta|=N}b_{\bm\beta}(-1)^{\alpha_M-\beta_M}\prod_{i=1}^{M-1}{\beta_i\choose \gamma_i},
\end{equation}
which is the desired relationship stated in Eqn. \eqref{eigenvector_components}.
\bibliography{Mvoter15}{}

\begin{thebibliography}{27}%
\makeatletter
\providecommand \@ifxundefined [1]{%
 \@ifx{#1\undefined}
}%
\providecommand \@ifnum [1]{%
 \ifnum #1\expandafter \@firstoftwo
 \else \expandafter \@secondoftwo
 \fi
}%
\providecommand \@ifx [1]{%
 \ifx #1\expandafter \@firstoftwo
 \else \expandafter \@secondoftwo
 \fi
}%
\providecommand \natexlab [1]{#1}%
\providecommand \enquote  [1]{``#1''}%
\providecommand \bibnamefont  [1]{#1}%
\providecommand \bibfnamefont [1]{#1}%
\providecommand \citenamefont [1]{#1}%
\providecommand \href@noop [0]{\@secondoftwo}%
\providecommand \href [0]{\begingroup \@sanitize@url \@href}%
\providecommand \@href[1]{\@@startlink{#1}\@@href}%
\providecommand \@@href[1]{\endgroup#1\@@endlink}%
\providecommand \@sanitize@url [0]{\catcode `\\12\catcode `\$12\catcode
  `\&12\catcode `\#12\catcode `\^12\catcode `\_12\catcode `\%12\relax}%
\providecommand \@@startlink[1]{}%
\providecommand \@@endlink[0]{}%
\providecommand \url  [0]{\begingroup\@sanitize@url \@url }%
\providecommand \@url [1]{\endgroup\@href {#1}{\urlprefix }}%
\providecommand \urlprefix  [0]{URL }%
\providecommand \Eprint [0]{\href }%
\providecommand \doibase [0]{http://dx.doi.org/}%
\providecommand \selectlanguage [0]{\@gobble}%
\providecommand \bibinfo  [0]{\@secondoftwo}%
\providecommand \bibfield  [0]{\@secondoftwo}%
\providecommand \translation [1]{[#1]}%
\providecommand \BibitemOpen [0]{}%
\providecommand \bibitemStop [0]{}%
\providecommand \bibitemNoStop [0]{.\EOS\space}%
\providecommand \EOS [0]{\spacefactor3000\relax}%
\providecommand \BibitemShut  [1]{\csname bibitem#1\endcsname}%
\let\auto@bib@innerbib\@empty
\bibitem [{\citenamefont {Liggett}(1999)}]{liggett}%
  \BibitemOpen
  \bibfield  {author} {\bibinfo {author} {\bibfnamefont {T.}~\bibnamefont
  {Liggett}},\ }\href@noop {} {\emph {\bibinfo {title} {Stochastic Interacting
  Systems: Contact, Voter, and Exclusion Processes}}}\ (\bibinfo  {publisher}
  {Springer-Verlag},\ \bibinfo {address} {New York},\ \bibinfo {year}
  {1999})\BibitemShut {NoStop}%
\bibitem [{\citenamefont {Liggett}(2005)}]{liggett2}%
  \BibitemOpen
  \bibfield  {author} {\bibinfo {author} {\bibfnamefont {T.~M.}\ \bibnamefont
  {Liggett}},\ }\href@noop {} {\emph {\bibinfo {title} {Interacting Particle
  Systems}}}\ (\bibinfo  {publisher} {Springer},\ \bibinfo {year}
  {2005})\BibitemShut {NoStop}%
\bibitem [{\citenamefont {Clifford}\ and\ \citenamefont
  {Sudbury}(1973)}]{clifford}%
  \BibitemOpen
  \bibfield  {author} {\bibinfo {author} {\bibfnamefont {P.}~\bibnamefont
  {Clifford}}\ and\ \bibinfo {author} {\bibfnamefont {A.}~\bibnamefont
  {Sudbury}},\ }\href@noop {} {\bibfield  {journal} {\bibinfo  {journal}
  {Biometrika}\ }\textbf {\bibinfo {volume} {60 (3)}},\ \bibinfo {pages}
  {581C588} (\bibinfo {year} {1973})}\BibitemShut {NoStop}%
\bibitem [{\citenamefont {Sen}\ and\ \citenamefont {Chakrabarti}(2013)}]{sen}%
  \BibitemOpen
  \bibfield  {author} {\bibinfo {author} {\bibfnamefont {P.}~\bibnamefont
  {Sen}}\ and\ \bibinfo {author} {\bibfnamefont {B.~K.}\ \bibnamefont
  {Chakrabarti}},\ }\href@noop {} {\emph {\bibinfo {title} {Sociophysics, An
  Introduction}}}\ (\bibinfo  {publisher} {Oxford University Press},\ \bibinfo
  {address} {Oxford},\ \bibinfo {year} {2013})\BibitemShut {NoStop}%
\bibitem [{\citenamefont {Castellano}\ \emph {et~al.}(2009)\citenamefont
  {Castellano}, \citenamefont {Fortunato},\ and\ \citenamefont
  {Loreto}}]{castellano}%
  \BibitemOpen
  \bibfield  {author} {\bibinfo {author} {\bibfnamefont {C.}~\bibnamefont
  {Castellano}}, \bibinfo {author} {\bibfnamefont {S.}~\bibnamefont
  {Fortunato}}, \ and\ \bibinfo {author} {\bibfnamefont {V.}~\bibnamefont
  {Loreto}},\ }\href@noop {} {\bibfield  {journal} {\bibinfo  {journal} {Rev.
  Mod. Phys.}\ }\textbf {\bibinfo {volume} {81}},\ \bibinfo {pages} {591}
  (\bibinfo {year} {2009})}\BibitemShut {NoStop}%
\bibitem [{\citenamefont {Starnini}\ \emph {et~al.}(2012)\citenamefont
  {Starnini}, \citenamefont {Baronchelli},\ and\ \citenamefont
  {Pastor-Satorras}}]{starnini}%
  \BibitemOpen
  \bibfield  {author} {\bibinfo {author} {\bibfnamefont {M.}~\bibnamefont
  {Starnini}}, \bibinfo {author} {\bibfnamefont {A.}~\bibnamefont
  {Baronchelli}}, \ and\ \bibinfo {author} {\bibfnamefont {R.}~\bibnamefont
  {Pastor-Satorras}},\ }\href@noop {} {\bibfield  {journal} {\bibinfo
  {journal} {J. Stat. Mech.}\ }\textbf {\bibinfo {volume} {2012}},\ \bibinfo
  {pages} {P10027} (\bibinfo {year} {2012})}\BibitemShut {NoStop}%
\bibitem [{\citenamefont {V{\'a}zquez}\ \emph {et~al.}(2003)\citenamefont
  {V{\'a}zquez}, \citenamefont {Krapivsky},\ and\ \citenamefont
  {Redner}}]{vazquez2}%
  \BibitemOpen
  \bibfield  {author} {\bibinfo {author} {\bibfnamefont {F.}~\bibnamefont
  {V{\'a}zquez}}, \bibinfo {author} {\bibfnamefont {P.}~\bibnamefont
  {Krapivsky}}, \ and\ \bibinfo {author} {\bibfnamefont {S.}~\bibnamefont
  {Redner}},\ }\href@noop {} {\bibfield  {journal} {\bibinfo  {journal} {J.
  Phys. A.}\ }\textbf {\bibinfo {volume} {36}},\ \bibinfo {pages} {L61}
  (\bibinfo {year} {2003})}\BibitemShut {NoStop}%
\bibitem [{\citenamefont {Volovik}\ \emph {et~al.}(2009)\citenamefont
  {Volovik}, \citenamefont {Mobilia},\ and\ \citenamefont {Redner}}]{volovik}%
  \BibitemOpen
  \bibfield  {author} {\bibinfo {author} {\bibfnamefont {D.}~\bibnamefont
  {Volovik}}, \bibinfo {author} {\bibfnamefont {M.}~\bibnamefont {Mobilia}}, \
  and\ \bibinfo {author} {\bibfnamefont {S.}~\bibnamefont {Redner}},\
  }\href@noop {} {\bibfield  {journal} {\bibinfo  {journal} {Europhys. Lett.}\
  }\textbf {\bibinfo {volume} {85}},\ \bibinfo {pages} {48003} (\bibinfo {year}
  {2009})}\BibitemShut {NoStop}%
\bibitem [{\citenamefont {Castell{\'o}}\ \emph {et~al.}(2006)\citenamefont
  {Castell{\'o}}, \citenamefont {Egu{\'iluz}},\ and\ \citenamefont
  {Miguel}}]{castello2}%
  \BibitemOpen
  \bibfield  {author} {\bibinfo {author} {\bibfnamefont {X.}~\bibnamefont
  {Castell{\'o}}}, \bibinfo {author} {\bibfnamefont {V.}~\bibnamefont
  {Egu{\'iluz}}}, \ and\ \bibinfo {author} {\bibfnamefont {M.~S.}\ \bibnamefont
  {Miguel}},\ }\href@noop {} {\bibfield  {journal} {\bibinfo  {journal} {New J.
  Phys.}\ }\textbf {\bibinfo {volume} {8}},\ \bibinfo {pages} {308} (\bibinfo
  {year} {2006})}\BibitemShut {NoStop}%
\bibitem [{\citenamefont {Baronchelli}\ \emph {et~al.}(2008)\citenamefont
  {Baronchelli}, \citenamefont {Loreto},\ and\ \citenamefont
  {Steels}}]{baronchelli2}%
  \BibitemOpen
  \bibfield  {author} {\bibinfo {author} {\bibfnamefont {A.}~\bibnamefont
  {Baronchelli}}, \bibinfo {author} {\bibfnamefont {V.}~\bibnamefont {Loreto}},
  \ and\ \bibinfo {author} {\bibfnamefont {L.}~\bibnamefont {Steels}},\
  }\href@noop {} {\bibfield  {journal} {\bibinfo  {journal} {Int. J. Mod. Phys.
  C.}\ }\textbf {\bibinfo {volume} {19}},\ \bibinfo {pages} {785} (\bibinfo
  {year} {2008})}\BibitemShut {NoStop}%
\bibitem [{\citenamefont {Zhang}\ \emph {et~al.}(2011)\citenamefont {Zhang},
  \citenamefont {Lim}, \citenamefont {Sreenivasan}, \citenamefont {Xie},
  \citenamefont {Szymanski},\ and\ \citenamefont {Korniss}}]{zhang}%
  \BibitemOpen
  \bibfield  {author} {\bibinfo {author} {\bibfnamefont {W.}~\bibnamefont
  {Zhang}}, \bibinfo {author} {\bibfnamefont {C.}~\bibnamefont {Lim}}, \bibinfo
  {author} {\bibfnamefont {S.}~\bibnamefont {Sreenivasan}}, \bibinfo {author}
  {\bibfnamefont {J.}~\bibnamefont {Xie}}, \bibinfo {author} {\bibfnamefont
  {B.}~\bibnamefont {Szymanski}}, \ and\ \bibinfo {author} {\bibfnamefont
  {G.}~\bibnamefont {Korniss}},\ }\href@noop {} {\bibfield  {journal} {\bibinfo
   {journal} {Chaos}\ }\textbf {\bibinfo {volume} {21}},\ \bibinfo {pages}
  {025115} (\bibinfo {year} {2011})}\BibitemShut {NoStop}%
\bibitem [{\citenamefont {Xie}\ \emph {et~al.}(2011)\citenamefont {Xie},
  \citenamefont {Sreenivasan}, \citenamefont {Korniss}, \citenamefont {Zhang},\
  and\ \citenamefont {Szymanski}}]{xie}%
  \BibitemOpen
  \bibfield  {author} {\bibinfo {author} {\bibfnamefont {J.}~\bibnamefont
  {Xie}}, \bibinfo {author} {\bibfnamefont {S.}~\bibnamefont {Sreenivasan}},
  \bibinfo {author} {\bibfnamefont {G.}~\bibnamefont {Korniss}}, \bibinfo
  {author} {\bibfnamefont {W.}~\bibnamefont {Zhang}}, \ and\ \bibinfo {author}
  {\bibfnamefont {B.}~\bibnamefont {Szymanski}},\ }\href@noop {} {\bibfield
  {journal} {\bibinfo  {journal} {Phys. Rev. E.}\ }\textbf {\bibinfo {volume}
  {84}},\ \bibinfo {pages} {011130} (\bibinfo {year} {2011})}\BibitemShut
  {NoStop}%
\bibitem [{\citenamefont {Lu}\ \emph {et~al.}(2009)\citenamefont {Lu},
  \citenamefont {Korniss},\ and\ \citenamefont {Szymanski}}]{lu}%
  \BibitemOpen
  \bibfield  {author} {\bibinfo {author} {\bibfnamefont {Q.}~\bibnamefont
  {Lu}}, \bibinfo {author} {\bibfnamefont {G.}~\bibnamefont {Korniss}}, \ and\
  \bibinfo {author} {\bibfnamefont {B.}~\bibnamefont {Szymanski}},\ }\href@noop
  {} {\bibfield  {journal} {\bibinfo  {journal} {J. Econ. Interact. Coord.}\
  }\textbf {\bibinfo {volume} {4}},\ \bibinfo {pages} {221} (\bibinfo {year}
  {2009})}\BibitemShut {NoStop}%
\bibitem [{\citenamefont {Lim}\ and\ \citenamefont {Pickering}(2014)}]{lim13}%
  \BibitemOpen
  \bibfield  {author} {\bibinfo {author} {\bibfnamefont {C.}~\bibnamefont
  {Lim}}\ and\ \bibinfo {author} {\bibfnamefont {W.}~\bibnamefont
  {Pickering}},\ }\href@noop {} {\bibfield  {journal} {\bibinfo  {journal}
  {arXiv:1411.0530, 2014}\ } (\bibinfo {year} {2014})}\BibitemShut {NoStop}%
\bibitem [{\citenamefont {Pickering}\ and\ \citenamefont
  {Lim}(2015)}]{pickering}%
  \BibitemOpen
  \bibfield  {author} {\bibinfo {author} {\bibfnamefont {W.}~\bibnamefont
  {Pickering}}\ and\ \bibinfo {author} {\bibfnamefont {C.}~\bibnamefont
  {Lim}},\ }\href@noop {} {\bibfield  {journal} {\bibinfo  {journal} {Phys.
  Rev. E}\ }\textbf {\bibinfo {volume} {91}},\ \bibinfo {pages} {012812}
  (\bibinfo {year} {2015})}\BibitemShut {NoStop}%
\bibitem [{\citenamefont {Kac}(1947)}]{kac}%
  \BibitemOpen
  \bibfield  {author} {\bibinfo {author} {\bibfnamefont {M.}~\bibnamefont
  {Kac}},\ }\href@noop {} {\bibfield  {journal} {\bibinfo  {journal} {Am. Math.
  Monthly}\ }\textbf {\bibinfo {volume} {54}},\ \bibinfo {pages} {369}
  (\bibinfo {year} {1947})}\BibitemShut {NoStop}%
\bibitem [{\citenamefont {John}(1981)}]{john}%
  \BibitemOpen
  \bibfield  {author} {\bibinfo {author} {\bibfnamefont {F.}~\bibnamefont
  {John}},\ }\href@noop {} {\emph {\bibinfo {title} {Partial Differential
  Equations}}}\ (\bibinfo  {publisher} {Springer-Verlag},\ \bibinfo {address}
  {New York},\ \bibinfo {year} {1981})\BibitemShut {NoStop}%
\bibitem [{\citenamefont {Newman}\ \emph {et~al.}(2001)\citenamefont {Newman},
  \citenamefont {Strogatz},\ and\ \citenamefont {Watts}}]{newman}%
  \BibitemOpen
  \bibfield  {author} {\bibinfo {author} {\bibfnamefont {M.}~\bibnamefont
  {Newman}}, \bibinfo {author} {\bibfnamefont {S.}~\bibnamefont {Strogatz}}, \
  and\ \bibinfo {author} {\bibfnamefont {D.}~\bibnamefont {Watts}},\
  }\href@noop {} {\bibfield  {journal} {\bibinfo  {journal} {Phys. Rev. E.}\
  }\textbf {\bibinfo {volume} {64}},\ \bibinfo {pages} {026118} (\bibinfo
  {year} {2001})}\BibitemShut {NoStop}%
\bibitem [{\citenamefont {Newman}(2002)}]{newman2}%
  \BibitemOpen
  \bibfield  {author} {\bibinfo {author} {\bibfnamefont {M.}~\bibnamefont
  {Newman}},\ }\href@noop {} {\bibfield  {journal} {\bibinfo  {journal} {Phys.
  Rev. E}\ }\textbf {\bibinfo {volume} {66}},\ \bibinfo {pages} {016128}
  (\bibinfo {year} {2002})}\BibitemShut {NoStop}%
\bibitem [{\citenamefont {Bender}\ and\ \citenamefont
  {Williamson}(2006)}]{bender}%
  \BibitemOpen
  \bibfield  {author} {\bibinfo {author} {\bibfnamefont {E.~A.}\ \bibnamefont
  {Bender}}\ and\ \bibinfo {author} {\bibfnamefont {S.~G.}\ \bibnamefont
  {Williamson}},\ }\href@noop {} {\emph {\bibinfo {title} {Foundations of
  Combinatorics with Applications}}}\ (\bibinfo  {publisher} {Dover},\ \bibinfo
  {address} {New York},\ \bibinfo {year} {2006})\BibitemShut {NoStop}%
\bibitem [{\citenamefont {Sood}\ and\ \citenamefont {Redner}(2005)}]{sood}%
  \BibitemOpen
  \bibfield  {author} {\bibinfo {author} {\bibfnamefont {V.}~\bibnamefont
  {Sood}}\ and\ \bibinfo {author} {\bibfnamefont {S.}~\bibnamefont {Redner}},\
  }\href@noop {} {\bibfield  {journal} {\bibinfo  {journal} {Phys. Rev. Lett.}\
  }\textbf {\bibinfo {volume} {94}},\ \bibinfo {pages} {178701} (\bibinfo
  {year} {2005})}\BibitemShut {NoStop}%
\bibitem [{\citenamefont {Waagen}\ \emph {et~al.}(2015)\citenamefont {Waagen},
  \citenamefont {Verma}, \citenamefont {Chan}, \citenamefont {Swami},\ and\
  \citenamefont {D'Souza}}]{waagen}%
  \BibitemOpen
  \bibfield  {author} {\bibinfo {author} {\bibfnamefont {A.}~\bibnamefont
  {Waagen}}, \bibinfo {author} {\bibfnamefont {G.}~\bibnamefont {Verma}},
  \bibinfo {author} {\bibfnamefont {K.}~\bibnamefont {Chan}}, \bibinfo {author}
  {\bibfnamefont {A.}~\bibnamefont {Swami}}, \ and\ \bibinfo {author}
  {\bibfnamefont {R.}~\bibnamefont {D'Souza}},\ }\href@noop {} {\bibfield
  {journal} {\bibinfo  {journal} {Phys. Rev. E}\ }\textbf {\bibinfo {volume}
  {91}},\ \bibinfo {pages} {022811} (\bibinfo {year} {2015})}\BibitemShut
  {NoStop}%
\bibitem [{\citenamefont {Castell{\'o}}\ \emph {et~al.}(2009)\citenamefont
  {Castell{\'o}}, \citenamefont {Baronchelli},\ and\ \citenamefont
  {Loreto}}]{castello}%
  \BibitemOpen
  \bibfield  {author} {\bibinfo {author} {\bibfnamefont {X.}~\bibnamefont
  {Castell{\'o}}}, \bibinfo {author} {\bibfnamefont {A.}~\bibnamefont
  {Baronchelli}}, \ and\ \bibinfo {author} {\bibfnamefont {V.}~\bibnamefont
  {Loreto}},\ }\href@noop {} {\bibfield  {journal} {\bibinfo  {journal} {Eur.
  Phys. J. B}\ }\textbf {\bibinfo {volume} {71}},\ \bibinfo {pages} {557}
  (\bibinfo {year} {2009})}\BibitemShut {NoStop}%
\bibitem [{\citenamefont {Zhang}(2012)}]{zhang_thesis}%
  \BibitemOpen
  \bibfield  {author} {\bibinfo {author} {\bibfnamefont {W.}~\bibnamefont
  {Zhang}},\ }\emph {\bibinfo {title} {Analytical Approach for Opinion Dynamics
  on Social Networks}},\ \href@noop {} {Ph.D. thesis},\ \bibinfo  {school}
  {Rensselaer Polytechnic Institute} (\bibinfo {year} {2012})\BibitemShut
  {NoStop}%
\bibitem [{\citenamefont {Ewens}(2004)}]{ewens}%
  \BibitemOpen
  \bibfield  {author} {\bibinfo {author} {\bibfnamefont {W.~J.}\ \bibnamefont
  {Ewens}},\ }\href@noop {} {\emph {\bibinfo {title} {Mathematical Population
  Genetics I. Theoretical Introduction}}},\ \bibinfo {edition} {2nd}\ ed.\
  (\bibinfo  {publisher} {Springer},\ \bibinfo {address} {New York},\ \bibinfo
  {year} {2004})\BibitemShut {NoStop}%
\bibitem [{\citenamefont {Moran}(1958)}]{moran}%
  \BibitemOpen
  \bibfield  {author} {\bibinfo {author} {\bibfnamefont {P.~A.~P.}\
  \bibnamefont {Moran}},\ }\href@noop {} {\bibfield  {journal} {\bibinfo
  {journal} {Mathematical Proceedings of the Cambridge Philosophical Society}\
  }\textbf {\bibinfo {volume} {54}},\ \bibinfo {pages} {60} (\bibinfo {year}
  {1958})}\BibitemShut {NoStop}%
\bibitem [{\citenamefont {Blythe}\ and\ \citenamefont {McKane}(2007)}]{blythe}%
  \BibitemOpen
  \bibfield  {author} {\bibinfo {author} {\bibfnamefont {R.~A.}\ \bibnamefont
  {Blythe}}\ and\ \bibinfo {author} {\bibfnamefont {A.~J.}\ \bibnamefont
  {McKane}},\ }\href@noop {} {\bibfield  {journal} {\bibinfo  {journal} {J.
  Stat. Mech.}\ }\textbf {\bibinfo {volume} {2007}},\ \bibinfo {pages} {P07018}
  (\bibinfo {year} {2007})}\BibitemShut {NoStop}%
\end{thebibliography}%

\end{document}